\documentclass[aip, pof,graphicx]{revtex4-2}

\usepackage{graphicx}
\usepackage{dcolumn}
\usepackage{bm}
\usepackage[utf8]{inputenc}
\usepackage[T1]{fontenc}
\usepackage{mathptmx}
\usepackage{color}
\usepackage{longtable}
\usepackage{epstopdf}

\draft 

\begin{document}


\title[Grain-scale computations of barchan dunes]{Grain-scale computations of barchan dunes\\
	This article may be downloaded for personal use only. Any other use requires prior permission of the author and AIP Publishing. This article appeared in Phys. Fluids 34, 123320 (2022) and may be found at https://doi.org/10.1063/5.0121810.} 



\author{Nicolao C. Lima}
\affiliation{School of Mechanical Engineering, UNICAMP - University of Campinas,\\
	Rua Mendeleyev, 200, Campinas, SP, Brazil
}

\author{Willian R. Assis}
\affiliation{School of Mechanical Engineering, UNICAMP - University of Campinas,\\
	Rua Mendeleyev, 200, Campinas, SP, Brazil
}

\author{Carlos A. Alvarez}
\altaffiliation[Present address: ]{Department of Geological Sciences, Stanford University, Stanford, CA 94305, USA}
\affiliation{Department of Atmospheric and Oceanic Sciences, University of California, Los Angeles,
Los Angeles, CA 90095-1565, USA
}

\author{Erick M. Franklin}%
 \email{erick.franklin@unicamp.br}
 \thanks{Corresponding author}
\affiliation{School of Mechanical Engineering, UNICAMP - University of Campinas,\\
	Rua Mendeleyev, 200, Campinas, SP, Brazil
}


\date{\today}

\begin{abstract}
Barchans are crescent-shaped dunes commonly found in diverse environments and scales: from the 10-cm-long barchans found under water to the 1-km-long barchans on Mars, passing by the 100-m-long dunes on Earth's deserts. Although ubiquitous in nature, there is a lack of grain-scale computations of the growth and evolution of those bedforms. In this paper, we investigate the values of grain properties (coefficients of sliding friction, rolling friction and restitution) necessary to carry out numerical simulations of subaqueous barchans with CFD-DEM (computational fluid dynamics - discrete element method), and how the values of those coefficients change the barchan dynamics. We made use of LES (large eddy simulation) for the fluid, varied the coefficients of sliding friction, rolling friction and restitution in the DEM, and compared the outputs with experiments. We show: (i) for the case of glass spheres, the values of coefficients for correctly obtaining the dune morphology, timescales, trajectories of individual grains, and forces experienced by grains; (ii) the LES meshes allowing computations of bedload while capturing the main disturbances of the fluid flow; (iii) how different values of coefficients affect the morphology of barchans; and (iv) that spheres with higher coefficients of rolling friction can be used for simulating barchans consisting of angular grains. Our results represent a significant step for performing simulations that capture, at the same time, details of the fluid flow (large eddies) and grains' motion (individual particles). 
\end{abstract}

\pacs{}

\maketitle 

\section{INTRODUCTION}
\label{sec:intro}

Sand dunes are basically waves or heaps over a granular bed, appearing by the action of a fluid flow. In all cases, the fluid flow must impose moderate shearing over the bed and form a moving granular layer that keeps contact with the fixed part of the bed, instead of suspending the grains. In the moving layer, called bedload, the grains move by rolling, sliding, small jumps, or ballistic flights \cite{Bagnold_1}, depending on the state and strength of the fluid flow (ballistic flights occur only in gases, the other three happening for both gases and liquids). When the fluid flows mainly in one direction and the amount of sediment is limited, dunes assume a crescent shape with horns pointing downstream, known as barchan dunes \cite{Bagnold_1, Andreotti_1, Charru_5, Courrech}. Barchans are found in diverse environments and scales: on Mars they can reach 1 km in length and take millennia to grow, on terrestrial deserts the scales are 100 m and years, while under water they have around 10 cm and grow in approximately one minute. Although ranging over several orders of magnitude, their morphology and dynamics are similar, showing that the crescent shape is a strong attractor.

There is a large number of barchan fields on Earth's deserts and on the surface of Mars, and the available data from remote sensing is increasing considerably. However, the large timescales of Martian and aeolian barchans hinder the assessment of their long-time evolution (in addition to a lack of grain-scale measurements in the Martian case). In a certain way, the same applies to detailed computations (at the grain scale) of barchan dunes, since computational costs are prohibitive given the large time and length scales of those barchans. On the other hand, the morphodynamics of subaqueous barchans is similar to those of their aeolian and Martian counterparts \cite{Hersen_1, Claudin_Andreotti, Parteli2} (with differences in grain trajectories and small differences in the morphology), but with much smaller time and length scales. Therefore, most of previous experiments \cite{Hersen_1, Endo, Franklin_8, Hori, Reffet, Alvarez, Alvarez3, Wenzel, Alvarez4, Alvarez6} and many of grain-scale simulations \cite{Alvarez5, Alvarez7} investigating the long-time evolution of barchans were conducted for the subaqueous case.

Typically, the experiments can be classified in those which investigated the morphology and celerity of dunes (measurements at the bedform scale) \cite{Hersen_1, Endo, Franklin_8, Hori, Reffet, Alvarez} and those which inquired into the motion of individual grains \cite{Alvarez3, Wenzel, Alvarez4, Assis2, Assis3} (measurements at the grain scale). While the former allowed the assessment of length and time scales, scaling laws, and celerities of barchans, the latter obtained trajectories and instantaneous velocities of moving grains, allowing the assessment of typical paths and velocities, local fluxes, transport rates, and mass gains/losses. Although valuable information was obtained from experiments using water tanks, some quantities are of difficult access, such as the instantaneous forces acting on each grain within a barchan dune (for the moment, it has not been measured experimentally).

As for the experiments, numerical simulations of barchans can be divided into those at the bedform and grain scales. The first numerical works were at the bedform scale and computed the evolution of aeolian barchans by using continuum models for the grains \cite{Sauermann_4, Herrmann_Sauermann, Kroy_A, Kroy_C, Kroy_B, Schwammle, Parteli4}. In those models, the fluid shear was computed directly by using analytical expressions for the perturbed flow \cite{Jackson_Hunt, Hunt_1, Weng}, and the grains supposed to move mainly in the longitudinal direction (with a small transverse diffusion), being entrained by saltation (ballistic flights). Important results were obtained from those simulations, which explained the instability mechanisms for the growth of aeolian barchans, their scales and length ratios, and the relation between the fluid flow and barchan morphology, for example. Khosronejad and Sotiropoulos \cite{Sotiropoulos, Khosronejad} used a different method for simulating subaqueous barchans: they coupled LES (large eddy simulation) for the water (and a water-suspension mixture) with a continuum model for the granular bed, so that they resolved the water flow at the scale of large eddies. Their simulations captured the main morphological characteristics of bedforms while an initially flat bed was deformed into a barchan field. Later, using the same method, Khosronejad et al. \cite{Khosronejad2} quantified the hydrodynamics mechanisms leading to the initiation and maturation of barchans and their interactions. Those simulations shed light on the morphodynamics of barchan dunes, but, because they used continuum models, did not capture the dynamics of grains. In addition, they are valid for barchans consisting of a large number of grains. When the barchans are small, the discrete nature of the granular material must be considered.

The first discrete computations of barchan dunes made use of simplified models such as the cellular automaton \cite{Narteau, Zhang_D}, where parcels representing ensembles of grains follow rules for their motion. By using a cellular automaton model also for the gas, Narteau et al. \cite{Narteau} carried out simulations of aeolian dunes, including an initial sand pile that was deformed into a barchan dune with superimposed bedforms. Their results, at the bedform scale, compared well with field data. With the same approach, Zhang et al. \cite{Zhang_D} simulated the growth and migration of aeolian dunes and obtained the mean residence time of sediments in steady-state barchans. Although their simulations were not at the grain scale, they were able to show that there is a balance between the divergent flux of grains (in the transverse direction) on the stoss slope and the convergent flux by avalanches on the lee side, from which they computed the residence time of grains and proposed that many of them remain trapped in a central slice of the barchan. Despite the increased level of detail gathered with the cellular automaton model, simulations at the grain scale are necessary in order to completely reproduce the grain-grain and grain-fluid interactions (and the resulting patterns).

To the best of our knowledge, the only numerical simulations of barchan dunes at the grain scale, and that consider the fluid flow at a similar scale, were those carried out by Alvarez and Franklin \cite{Alvarez5, Alvarez7} for subaqueous dunes. Alvarez and Franklin \cite{Alvarez5} simulated the formation of barchan dunes by a turbulent water flow by coupling DEM (discrete element method) with LES. In the simulations, a pile consisting of 40,000 glass spheres was deformed into a barchan dune by the water flow, and the instantaneous flow field for the water, and positions, velocities and forces for each grain were accessible. The use of LES allowed computations of the water flow at the grain scale (the LES grid and the particle diameter were of the same order) at lower computational cost than DNS (direct numerical simulation). As main results, the same morphodynamics observed in previous experiments \cite{Alvarez, Alvarez3, Alvarez4} were obtained at both the barchan and grain scales, such as the typical lengths and celerities of barchans, and typical paths and velocities of grains. Later, using the same LES-DEM simulations, Alvarez and Franklin \cite{Alvarez7} inquired into the forces experienced by each grain within a barchan dune, showing how forces are distributed and transmitted in a barchan. They showed that the distribution of the resultant force tends to route a significant part of grains toward the crest and horns and, in particular, that the longitudinal component of the resultant force on a great part of grains is of the order of 10$^{-7}$ N, with negative values around the crest (which results in grain deposition on the crest). Their results revealed the process, at the grain scale, for reaching the crescent shape of barchans.

Although ubiquitous in nature, there are few numerical computations of the growth and evolution of barchans carried out at the grain scale. As a result, values of solid-solid coefficients and size restrictions of CFD (computational fluid dynamics) meshes have not been exhaustively investigated for this specific problem. In this paper, we investigate the values of the coefficients of sliding friction, rolling friction and restitution necessary to carry out numerical simulations of subaqueous barchans with CFD-DEM (computational fluid dynamics - discrete element method), and how the values of those coefficients change the barchan dynamics. We use LES to compute the water flow at a scale similar to that of grains, vary the coefficients of sliding friction, rolling friction and restitution in the DEM, and compare the numerical outputs with experiments conducted under the same conditions.  We show: (i) for the case of glass spheres, the values of coefficients for correctly obtaining the dune morphology, timescales, trajectories of individual grains, and forces experienced by grains; (ii) the LES meshes allowing computations of bedload while capturing the main disturbances of the fluid flow; (iii) how different values of coefficients affect the morphology of barchans; and (iv) that spheres with higher coefficients of rolling friction can be used for simulating barchans consisting of angular grains. Our numerical results compare well with experiments, representing a significant step for performing simulations that capture, at the same time, details of the fluid flow (large eddies) and grains' motion (individual particles).

In the following, Secs. \ref{sec:formulation} and \ref{sec:methods} present, respectively, the fundamental and implemented equations, Sec.  \ref{sec:setup} shows the numerical setup, and Sec. \ref{sec:results} presents the results for the water and grains. Section \ref{sec:conclusions} concludes the paper.

\section{MATHEMATICAL FORMULATION}
\label{sec:formulation}

In CFD-DEM computations, solid particles are followed in a Lagrangian framework while the fluid flow is computed in an Eulerian grid. The basic equations consist, therefore, of Newton's law of motion, for both the grains and fluid. For the grains, the linear (Eq. \ref{Fp}) and angular momentum (Eq. \ref{Tp}) equations are computed in a Lagrangian framework,

\begin{equation}
	m_{p}\frac{d\vec{u}_{p}}{dt}= \vec{F}_{fp} + \vec{F}_{c} + m_{p}\vec{g}\, ,
	\label{Fp}
\end{equation}

\begin{equation}
	I_{p}\frac{d\vec{\omega}_{p}}{dt}=\vec{T}_{c} 
+ \vec{T}_{f}\, ,
	\label{Tp}
\end{equation}

\noindent where $\vec{g}$ is the acceleration of gravity and, for each grain, $m_{p}$ is the mass, $\vec{u}_{p}$ is the velocity, $\vec{F}_{fp}$ is the resultant of fluid forces, $\vec{F}_{c}$ is the resultant of contact forces between solids, $I_{p}$ is the moment of inertia, $\vec{\omega}_{p}$ is the angular velocity, $\vec{T}_{c}$ is the resultant of contact torques between solids, and $\vec{T}_{f}$ is the resultant of torques caused by the fluid. The resultant of fluid forces on a grain is the sum of the forces caused by fluid drag $\vec{F}_{d}$, pressure gradient $\vec{F}_{p}$, deviatoric stress tensor $\vec{F}_{\tau}$, virtual mass $\vec{F}_{vm}$, Magnus effect $\vec{F}_{Mag}$, Saffman effect $\vec{F}_{Saff}$, and Basset $\vec{F}_{Bass}$.

The contact forces and torques are computed by Eqs. \ref{Fc} and \ref{Tc}, respectively,

\begin{equation}
	\vec{F}_{c} = \sum_{i\neq j}^{N_c} \left(\vec{F}_{c,ij} \right) + \sum_{i}^{N_w} \left( \vec{F}_{c,iw} \right)
	\label{Fc}
\end{equation}

\begin{equation}
	\vec{T}_{c} = \sum_{i\neq j}^{N_c} \vec{T}_{c,ij} + \sum_{i}^{N_w} \vec{T}_{c,iw}
	\label{Tc}
\end{equation}

\noindent where $\vec{F}_{c,ij}$ and $\vec{F}_{c,iw}$ are the contact forces between particles $i$ and $j$ and between particle $i$ and the wall, respectively, $\vec{T}_{c,ij}$ is the torque due to the tangential component of the contact force between particles $i$ and $j$, $\vec{T}_{c,iw}$ is the torque due to the tangential component of the contact force between particle $i$ and the wall, $N_c$ - 1 is the number of particles in contact with particle $i$, and $N_w$ is the number of particles in contact with the wall. More details on the forces and torques are described in Appendix \ref{appendix}.

For the flow of a liquid, the incompressible mass (Eq. \ref{mass}) and momentum (Eq. \ref{mom}) equations are computed in an Eulerian framework, 

\begin{equation}
	\nabla\cdot\vec{u}_{f}=0 \, ,
	\label{mass}
\end{equation}

\begin{equation}
	\frac{\partial \left(\rho_{f}\vec{u}_{f} \right) }{\partial{t}} + \nabla \cdot (\rho_{f}\vec{u}_{f}\vec{u}_{f}) = -\nabla P + \nabla\cdot \vec{\vec{\tau}} + \rho_{f}\vec{g} - \frac{N}{V}\vec{F}_{fp} \, ,
	\label{mom}
\end{equation}

\noindent where $\vec{u}_{f}$ is the fluid velocity, $\rho_{f}$ is the fluid density, $N$ is the number of grains and $V$ is the considered volume.

\section{IMPLEMENTED EQUATIONS}
\label{sec:methods}

We made use of the open-source code \mbox{CFDEM} \cite{Goniva} (www.cfdem.com), which couples the open-source CFD code OpenFOAM with the open-source DEM code LIGGGHTS \cite{Kloss, Berger}, for carrying out our LES-DEM computations. We describe briefly in the following how the fundamental equations presented in Section \ref{sec:formulation} are implemented in those codes, a detailed description of contact and force implementations being found in Appendices \ref{appendix} and \ref{appendix2}, as well as in Goniva et al. \cite{Goniva}.

The DEM code computes the Lagrangian motion of grains by using the linear and angular momentum equations, Eqs. \ref{Fp} and \ref{Tp}, respectively. In the resultant of fluid forces acting on a particle, $\vec{F}_{fp}$, we neglect the Basset, Saffman and Magnus forces, since they are usually considered of lesser importance for CFD-DEM simulations \cite{Zhou}. Therefore, our DEM simulations compute

\begin{equation}
	\vec{F}_{fp} = \vec{F}_{d} + \vec{F}_{p} + \vec{F}_{\tau} + \vec{F}_{vm} \, ,
	\label{Ffp_sim}
\end{equation}

\noindent where the Archimedes force is considered in $\vec{F}_{p}$ (although called explicitly in the numerical code, as explained next). In the angular momentum, Eq. \ref{Tp}, we neglect torques caused by the fluid because those due to contacts are usually much higher \cite{Tsuji, Tsuji2, Liu}. For the contact forces and torques, Eqs. \ref{Fc} and \ref{Tc}, we consider a Hertzian model. In this model, $\vec{F}_{c,ij}$ and $\vec{F}_{c,iw}$ (Eq. \ref{Fc}), are decomposed into normal and tangential components, given by Eqs. \ref{eqnFcn} and \ref{eqnFct},

\begin{equation}
	F_{c,n} = \kappa_{n}\delta_{n} - \gamma_{n}\frac{d\delta_{n}}{dt} \, ,
	\label{eqnFcn}
\end{equation}

\begin{equation}
	F_{c,t} = \kappa_{t}\delta_{t} - \gamma_{t}\frac{d\delta_{t}}{dt} \, ,
	\label{eqnFct}
\end{equation}

\noindent where the two terms on the RHS of Eq. \ref{eqnFcn} correspond to a repulsive force and a viscoelastic damping, $\delta_{n}$ being the normal displacement of two solids in contact, and $\kappa_{n}$ and $\gamma_{n}$ coefficients described in Appendix \ref{appendix} (note that $\kappa_{n}$ $\sim$ $\delta_{n}^{1/2}$, so that $F_{c,n}$ $\sim$ $\delta_{n}^{3/2}$). The two terms on the RHS of Eq. \ref{eqnFct} correspond to a shear force and a viscoelastic damping, $\delta_{t}$ being the tangential displacement measured in the direction perpendicular to the plane of contact, and $\kappa_{t}$ and $\gamma_{t}$ coefficients also described in Appendix \ref{appendix}. $F_{c,t}$ is computed by Eq. \ref{eqnFct}, but it is limited by the Coulomb's Law, 

\begin{equation}
	F_{c,t} = \mu F_{c,n}\, ,
	\label{eqncoulomb}
\end{equation}

\noindent where $\mu$ is the microscopic coefficient of sliding friction. Once that limit is reached, $F_{c,t}$ is computed by Eq. \ref{eqncoulomb} until the contact ends. In our simulations, we use a single value for $\mu$ (we do not differentiate between static and dynamic coefficients).  More details about the contact model are described in Appendix \ref{appendix}.

The CFD part computes the dynamics of the fluid phase in an Eulerian framework by solving the incompressible mass (Eq. \ref{mass}) and momentum (Eq. \ref{mom}) equations. Given the large number of solid particles in the simulations, we use an unresolved approach where the incompressible flow equations are phase-averaged (volume basis) while assuring mass conservation. In the code, they correspond to the phase-averaged mass and momentum equations of \textit{Set II} of Zhou et al. \cite{Zhou}, given by Eqs. \ref{mass_fluid} and \ref{momentum_fluid}, respectively.

\begin{equation}
	\frac{\partial \left( \alpha_{f} \rho_{f} \right)}{\partial t} + \nabla \cdot \left ( \alpha_{f} \rho_{f} \vec{u}_{f} \right ) = 0
	\label{mass_fluid}
\end{equation}

\begin{equation}
	\frac{\partial \left ( \alpha_{f} \rho_{f} \vec{u}_{f} \right ) }{\partial t} + \nabla \cdot \left ( \alpha_{f} \rho_{f} \vec{u}_{f} \vec{u}_{f} \right ) = -\alpha_{f} \nabla P - \vec{f}_{exch} + \alpha_{f} \nabla \cdot  \vec{\vec{\tau}}_{f}  + \alpha_{f} \rho_{f} \vec{g}
	\label{momentum_fluid}
\end{equation}

\noindent where $\alpha_{f}$ is the volume fraction of the fluid. In the averaging process, the forces due to the pressure gradient and deviatoric stress tensor were separated from the remaining fluid-particle forces, so that those terms appear explicitly in Eq. \ref{momentum_fluid}. The remaining phase-averaged forces (per unit of volume) that the fluid applies on solid particles, $\vec{f}_{exch}$, are given by Eq. \ref{forces_exchange} (since we neglected Basset, Saffman and Magnus forces),

\begin{equation}
	\vec{f}_{exch} = \frac{1}{\Delta V}\sum_{i}^{n_{p}} \left( \vec{F}_{d} +  \vec{F}_{vm} \right) \, ,
	\label{forces_exchange}
\end{equation}

\noindent where $n_p$ is the number of particles in the considered cell, whose volume is $\Delta V$. More details on the computation of $\vec{f}_{exch}$ are available in Appendix \ref{appendix2}.

Note: since the gravitational term of Eq. \ref{momentum_fluid} is absent in the PISO (pressure-implicit with splitting of operators) algorithm and $P$ represents the static pressure, the Archimedes force is explicitly called by the numerical code when computing Eq. \ref{Ffp_sim}.

\section{NUMERICAL SETUP}
\label{sec:setup}

The choice of an appropriate model depends mainly on how the scale of solid particles compares with the one being solved for the fluid flow. In our case, the grain diameter is small enough to fit in one computational cell, and therefore the unresolved formulation can be used. For the CFD, we use LES with the wall-adapting local eddy-viscosity (WALE) model \cite{Nicoud} in order to capture all re-circulations and high-energy vortices. However, since the mesh size in LES must be small enough to capture the desired scales of the flow, locating the solid particles using the position of their centroids can lead to discontinuities and instabilities (this has been reported to happen when $\Delta x < 3d$, where $d$ is the grain diameter and $\Delta x$ the grid size \cite{Pirker, Marshall}). Therefore, we use the \textit{divided} approach implemented in CFDEM, which divides each solid particle into 29 regions of same volume \cite{kloss2}. Each region is represented by its center point, so that different regions of a grain can be in different mesh elements, smoothing the void fraction and the momentum-exchange term, and allowing computations with grains whose sizes are comparable to those of mesh elements.  Besides the \textit{divided} approach, we also use a diffusion equation in order to smooth even more the void fraction, making the fluid-solid coupling more stable \cite{Pirker, Blais, Delacroix} (more details in Appendix \ref{appendix3}).

The fluid domain is a channel of size $L_x$ = 0.4 m, $L_y$ = $\delta$ = 0.025 m and $L_z$ = 0.1 m, where $x$, $y$ and $z$ are the longitudinal, vertical and spanwise directions, respectively, with periodic conditions in the longitudinal and spanwise directions. In order to save computing time, the vertical dimension of the domain, $L_y$ = $\delta$, corresponds to the channel half height (the upper limit in the $y$ direction corresponds to the channel centerline, the real channel height being 2$\delta$). The fluid is water, flowing with a cross-sectional mean velocity $U$ = 0.28 m/s. The channel Reynolds number based on $U$, Re = $U 2\delta \nu^{-1}$, is 14,000 and that based on the shear velocity $u_*$, Re$_*$ = $u_* \delta \nu^{-1}$, is 400, where $\nu$ is the kinematic viscosity (10$^{-6}$ m$^2$/s for water). We tested convergence by using different meshes, that are shown in Tab. \ref{meshProps}, where $d_x$, $d_y$ and $d_z$ are the numbers of divisions of the domain in the $x$, $y$ and $z$ directions, respectively, and the column $y^{+}_{1st}$ corresponds to the vertical position of the center of the first control volume (in the wall-normal direction and scaled in inner-wall units, $y^{+}$ = $y u_*/\nu$). To reduce computational costs while still maintaining accurate results, we used the mesh of type 1 (Tab. \ref{meshProps}) in our simulations (the convergence tests are shown in Sec. \ref{sec:results}).

\begin{table}[!h]
	\centering
	\caption{For each mesh type, the numbers of control volumes $N_{cv}$ and of divisions of the domain in the $x$, $y$ and $z$ directions ($d_x$, $d_y$ and $d_z$, respectively), and the vertical position of the center of the first control volume scaled in inner-wall units, $y^{+}_{1st}$.}
	\begin{tabular}{cccccc}
		\hline
		Mesh & $N_{cv}$ & $d_x$ & $d_y$ & $d_z$ & $y^{+}_{1st}$ \\ \hline
		1    & 472500          & 250   & 30    & 63    & 2.1     \\
		2    & 708750          & 250   & 45    & 63    & 1.4     \\
		3    & 945000          & 250   & 60    & 63    & 1.0     \\ \hline
	\end{tabular}
	\label{meshProps}
\end{table}

\begin{figure}[h!]
	\centering
	\includegraphics[width=0.9\columnwidth]{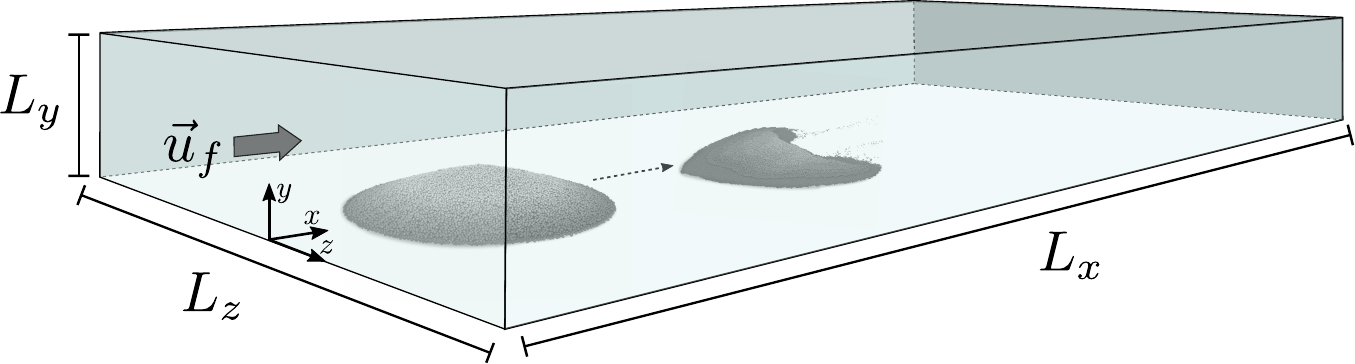}
	\caption{Layout of the numerical setup. $\vec{u_f}$ represents the water velocity, and $L_x$ = 0.4 m, $L_y$ = 0.025 m and $L_z$ = 0.1 m  are the dimensions of the computational domain in the $x$, $y$ and $z$ directions. The layout shows the initial pile and, at a later time, the barchan dune (the initial pile is deformed into the barchan dune, as indicated by the arrow). Multimedia view.}
	\label{Setup}
\end{figure}

Before starting the CFD-DEM simulations, a water flow in the absence of particles is simulated in the periodic channel until fully-developed turbulence is achieved. Afterward, the results are stored and a new computation starts in which the solid spheres (grains) are allowed to fall freely in stationary water, forming a conical heap with radius $R$ $\approx$ 0.013 m, height $h$ $\approx$ 0.003 m (values vary depending on the friction coefficients), and centered $4$ cm downstream from the channel inlet. Finally, CFD-DEM the simulations begin by imposing the previously stored turbulent flow, which deforms the conical pile into a barchan dune. Figure \ref{Setup} (Multimedia view) shows the setup for a typical simulation.

\begin{table}[!h]
	\centering
	\caption{Distribution of diameters for the grains in the initial pile, where $N_{d}$ is the number of grains for each diameter $d$, and physical properties.}
	\begin{tabular}{cc|lc}
		\hline
		$N_{d}$ & Diameter $d$ ($mm$) & \multicolumn{2}{l}{Properties} \\ \hline
		2276       & 0.15  &   Sliding Friction Coeff. $\mu$   & 0.1 to 0.9   \\
		13592      & 0.175  &  Rolling Friction Coeff. $\mu_r$  & 0.01 to 0.3   \\
		68267      & 0.2    &  Restitution Coef. $e$   & 0.1 to 0.9  \\
		13591      & 0.225  &  Poisson Ratio $\sigma$   & 0.45   \\
		2274       & 0.25   &  Young's Modulus $E$ (MPa) & 5    \\
		$\cdots$       & $\cdots$   &  Density $\rho_p$ (kg/m$^3$)  & 2500  \\ \hline
		
	\end{tabular}
	\label{diameterAndProps}
\end{table}

The grains consisted of  10$^5$ glass spheres randomly distributed, with sizes following a Gaussian distribution within 0.15 mm $\leq$ $d$ $\leq$ 0.25 mm, and we varied their physical properties within large ranges. The distribution of particle diameters, the ranges of the coefficients of sliding friction $\mu$, rolling friction $\mu_r$ and restitution $e$, and the values of Poisson ratio $\sigma$, Young's modulus $E$ and density $\rho_p$ used in the simulations are shown in Tab. \ref{diameterAndProps}. We note that we varied each property ($\mu$, $\mu_r$ or $e$) independently, keeping two of the following coefficients at $\mu$ = 0.6, $\mu_r$ = 0.0 and $e$ = 0.1, while the third one varied within the values in Tab. \ref{diameterAndProps}.

For the fluid, the boundary conditions were periodic flow in the longitudinal and transverse directions, impermeability and no-slip conditions at the bottom wall, and free slip on the top boundary ($y$ = $\delta$). For the grains, the boundary conditions were solid wall at the bottom boundary and free exit at the outlet. No influx of grains was imposed, so that the bedform lost grains and decreased slightly in size along time, in the same manner as in our previous experiments \cite{Alvarez, Alvarez3, Alvarez4, Alvarez6}. The time step used for the DEM was 2.5 $\times$ 10$^{-6}$ s, which was less than 20 \% of the Rayleigh time \cite{Derakhshani} in all simulated cases, and that of CFD was 2.5 $\times$ 10$^{-4}$ s, which respected the CFL (Courant-Friedrichs-Lewy) criterion \cite{Courant} (see the supplementary material for graphics showing convergence parameters). The numerical scripts are available in a data repository \cite{Supplemental3}.

\section{RESULTS}
\label{sec:results}

\subsection{LES convergence test}

\begin{figure}[h]
	\centering
	\includegraphics[width=0.95\columnwidth]{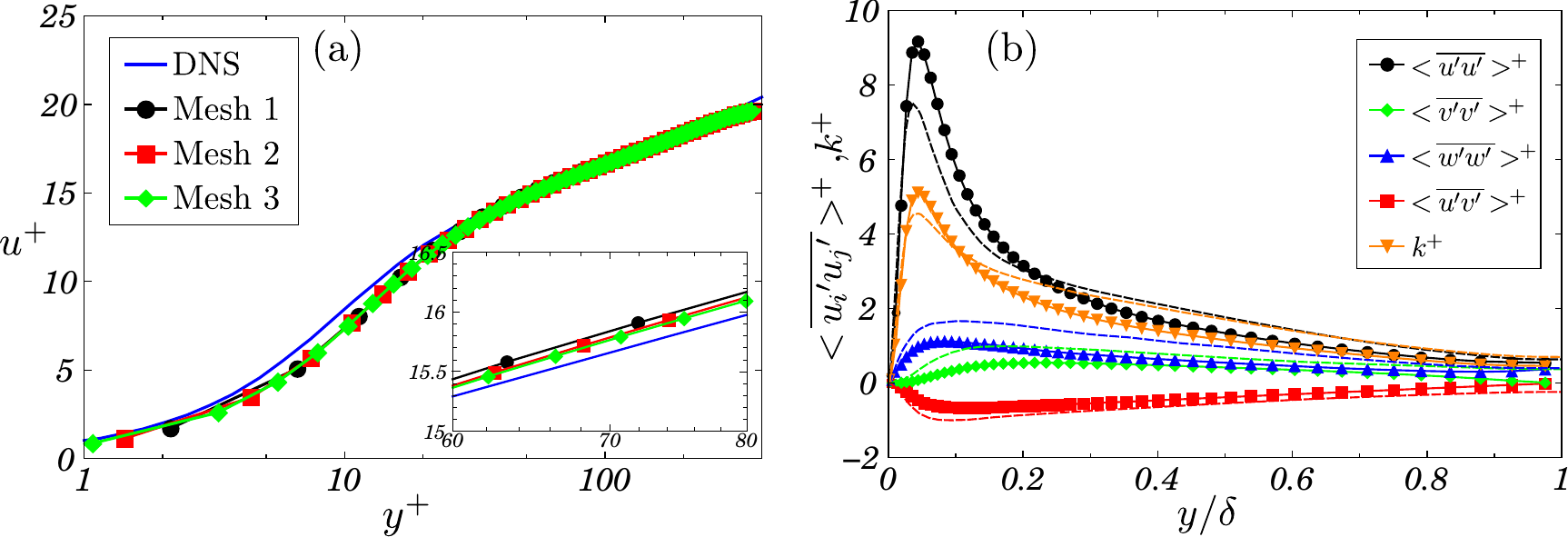}
	\caption{(a) For each mesh, profile of the longitudinal component of the mean velocity in log-normal scales, normalized by the inner scales. (b) For mesh 1, profiles of the turbulent kinetic energy and components of the Reynolds stress normalized by the inner scales. Blue line in figure (a) and dashed lines in figure (b) represent the DNS results of Moser et al. \cite{Moser}.}
	\label{MeshConvergence}
\end{figure}

We start by performing an analysis of mesh convergence for the LES computations of a single-phase water flow in the channel. The three different computational meshes are shown in Tab. \ref{meshProps}, and the results are compared with the DNS results of Moser et al. \cite{Moser} in Fig. \ref{MeshConvergence}a. For the different meshes, this figure shows the profile of the longitudinal component of the time-averaged velocity $\overline{u}$ normalized by the inner scales, i.e., $u^+$ = $\overline{u}/u_*$ as a function of $y^+$. From the inset of Fig. \ref{MeshConvergence}a, we observe the convergence of results as the mesh gets thinner, with very small deviations between results of meshes 1 to 3. For mesh 1, Fig. \ref{MeshConvergence}b presents the profiles of the turbulent kinetic energy and components of the Reynolds stress, where the superscript $+$ stands for normalization by the inner scales (division by the square of $u_{*}$), $<>$ means space averages in the $x$ and $z$ directions, and overbar refers to time averages. We consider that the agreement with the DNS results of Moser et al. \cite{Moser} is good for the three mesh sizes, and, since mesh 1 has almost half of the computational cost of mesh 3, we performed all the following computations using mesh 1.

\subsection{Flow over barchans}

\begin{figure}[h!]
	\centering
	\includegraphics[width=0.95\columnwidth]{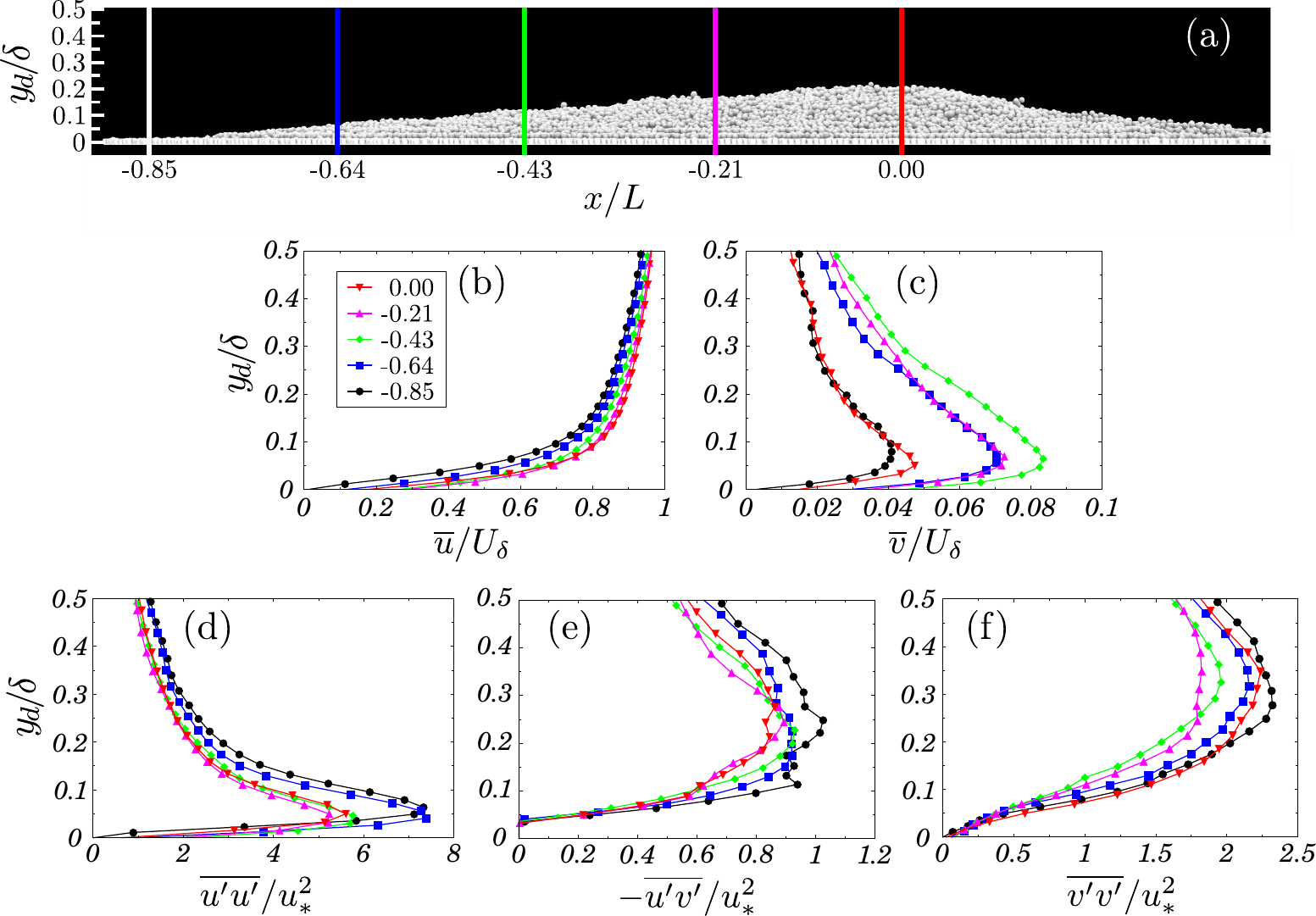}
	\caption{Water flow over a barchan dune, computed at the vertical symmetry plane. (a) Longitudinal positions, normalized by the barchan length $L$ (along the barcan centerline), of the profiles shown next. (b-f) Vertical profiles of (b) longitudinal $\overline{u}$ and (c) vertical $\overline{v}$ mean velocities, and second order moments (d) $\overline{u'u'}$, (e) $\overline{v'v'}$ and (f) $-\overline{u'v'}$, at the longitudinal positions previously indicated. The displaced coordinate $y_d$ is normalized by the channel half height $\delta$, the mean velocities by the mean velocity at the channel centerline $U_{\delta}$, and the second order moments by $u_*^2$. Profiles are in streamline coordinates.}
	\label{fluid_over_dune}
\end{figure}

For the simulations with grains, we measured the water flow over a barchan dune. For that, we carried out one specific simulation in which we let the pile be deformed into a barchan dune, froze the barchan from a given time on, and stored the velocity fields during 100 s. We then time averaged the flow and obtained the mean velocities and second order moments. Figure \ref{fluid_over_dune}b shows the vertical profiles of $\overline{u}$, $\overline{v}$, $\overline{u'u'}$, $\overline{v'v'}$ and $-\overline{u'v'}$ in streamline coordinates, each one at five different longitudinal positions of the barchan dune as shown in Fig. \ref{fluid_over_dune}a. The profiles were computed in the vertical plane of symmetry of the barchan, and they use the displaced coordinate (Eq. \ref{eq_yd}), i.e., the vertical distance from the dune surface, 

\begin{equation}
	y_d = y - h(x) \, ,
	\label{eq_yd}
\end{equation}

\noindent where $h(x)$ is the local height of the barchan along its symmetry line. In Figs. \ref{fluid_over_dune}b-f, $y_d$ is normalized by the channel half height $\delta$, $\overline{u}$ and $\overline{v}$ by the mean velocity at the channel centerline $U_{\delta}$, and the second order moments by $u_*^2$.

We observe that the results are very close to those of Charru and Franklin \cite{Franklin_11}, who carried out experiments in which turbulent flows over a barchan dune were measured with PIV (particle image velocimetry). The agreement between our results and those in Ref. \cite{Franklin_11} is excellent for $\overline{u}$, $\overline{v}$ and $-\overline{u'v'}$, while that for $\overline{u'u'}$ and $\overline{v'v'}$ is qualitatively good (the numerical results have higher values). We presume that the discrepancies in magnitude for $\overline{u'u'}$ and $\overline{v'v'}$ are due to noise and filtering in the PIV experiments (caused by undesirable reflections) and small differences in the shapes of the barchans (numerical vs. experimental). In any case, however, the  $-\overline{u'v'}$ component is the most relevant for the transport of grains along the dune. 

Part of our LES results can be also compared with those of Khosronejad et al. \cite{Khosronejad2}. Because in their experiments and LES computations the barchan dune was much larger than ours (around 1 m in length) and covered with ripples (with size comparable to our barchans), we examine here only $\overline{u}$ and $-\overline{u'v'}$ over the dune. Comparing thus the vertical profiles over the dune presented by Khosronejad et al. \cite{Khosronejad2} in the crest region with our results, we notice a good agreement in terms of format and absolute values (noting that their graphics are for $-\overline{u'v'}/U^2$, with $u_*$ $=$ 0.042$U$).

Therefore, we consider that the water flow over the barchan is well resolved with the LES approach. We also measured the flow over a barchan with moving grains (i.e., the barchan was not frozen) and the differences are negligible (see the supplementary material for the flow profiles over a barchan with the presence of bedload).

\subsection{Dune morphology}

\begin{figure}[h!]
	\centering
	\includegraphics[width=0.95\columnwidth]{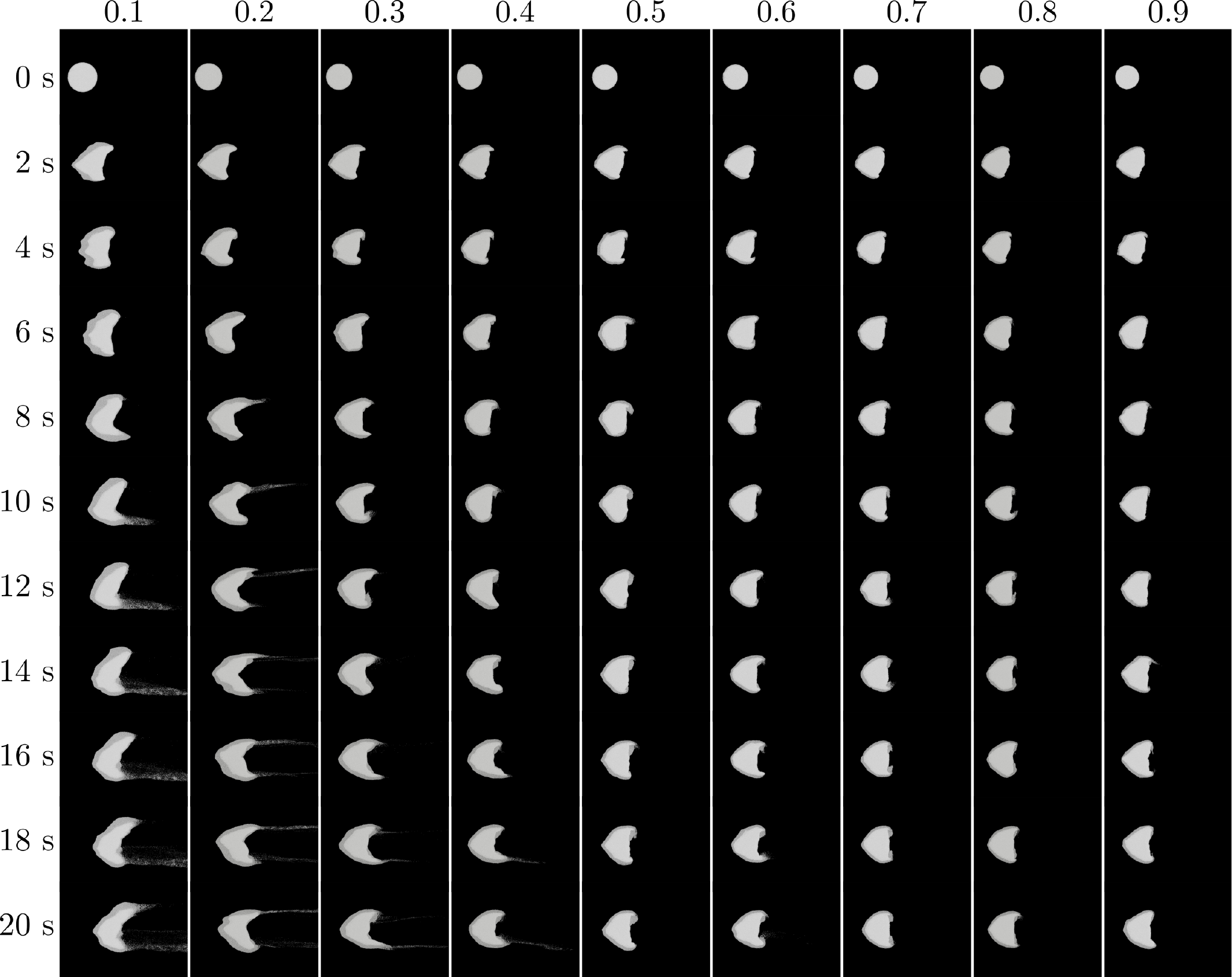}
	\caption{Snapshots showing top views of grains for different coefficients of sliding friction $\mu$. Each column corresponds to a friction coefficient (values shown on the top), and each row to a time instant (values shown on the left). In this figure, $\mu_r$ = 0.0 and $e$ = 0.1.}
	\label{morphology1}
\end{figure}

\begin{figure}[h!]
	\centering
	\includegraphics[width=0.55\columnwidth]{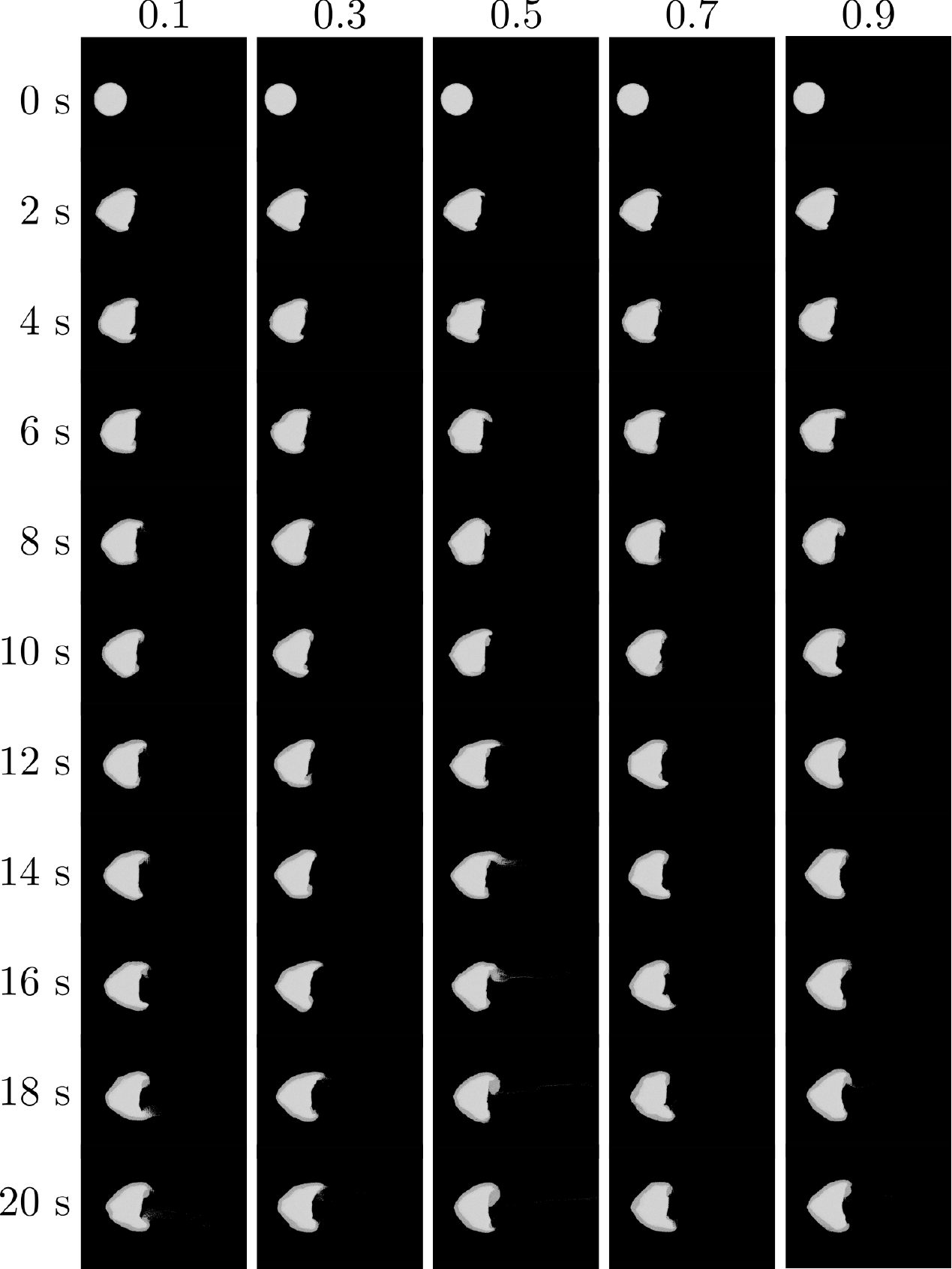}
	\caption{Snapshots showing top views of grains for different restitution coefficients $e$. Each column corresponds to a restitution coefficient (values shown on the top), and each row to a time instant (values shown on the left). In this figure, $\mu$ = 0.6 and $\mu_r$ = 0.0.}
	\label{morphology2}
\end{figure}

\begin{figure}[h!]
	\centering
	\includegraphics[width=0.55\columnwidth]{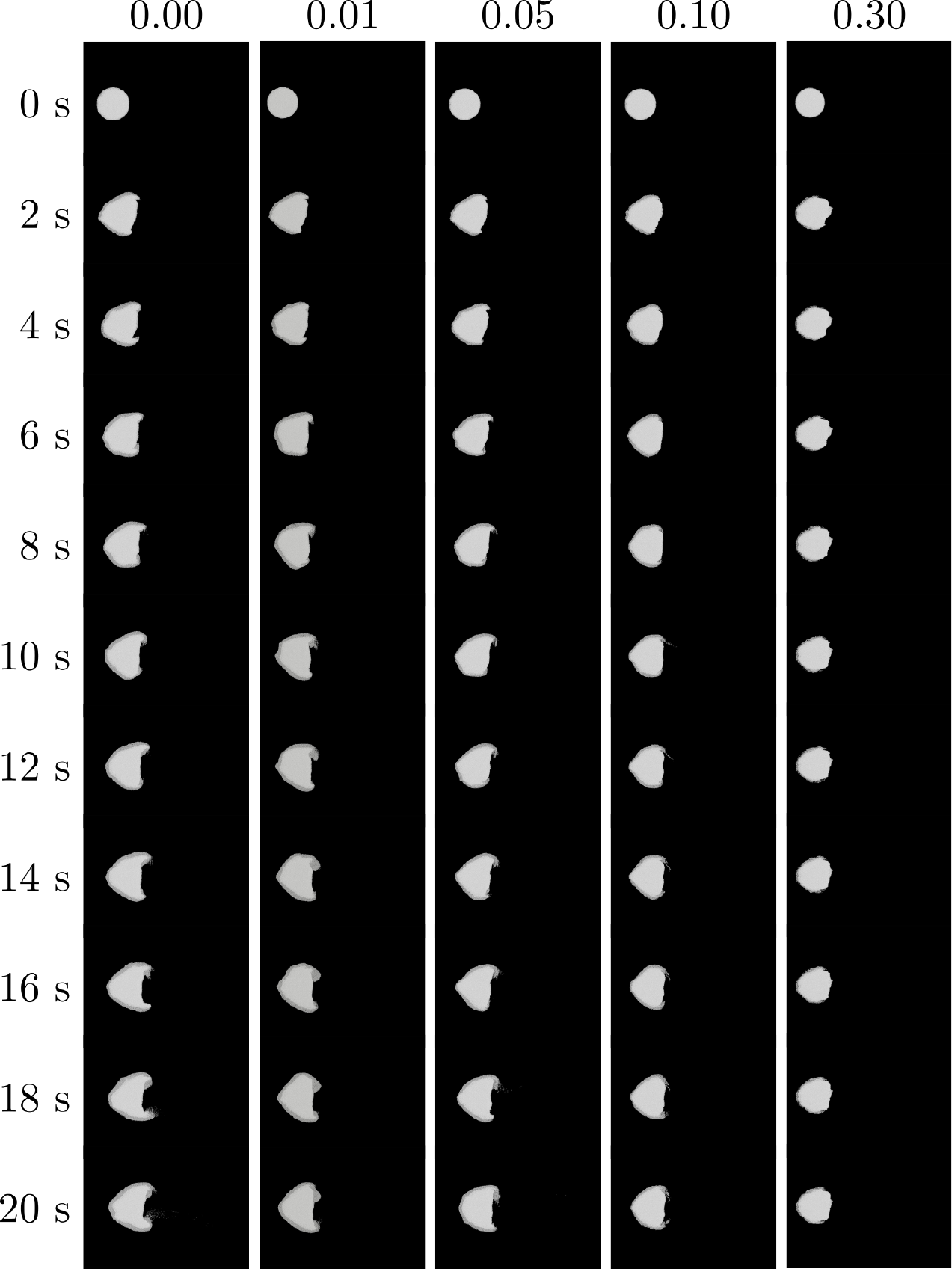}
	\caption{Snapshots showing top views of grains for different coefficients of rolling friction $\mu_r$. Each column corresponds to a coefficient of rolling friction (values shown on the top), and each row to a time instant (values shown on the left). In this figure, $\mu$ = 0.6 and $e$ = 0.1.}
	\label{morphology3}
\end{figure}

\begin{figure}[h!]
	\centering
	\includegraphics[width=0.8\columnwidth]{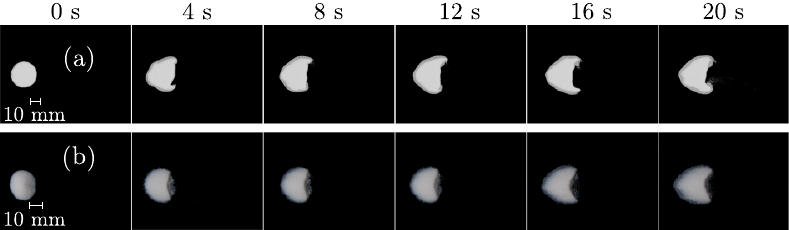}
	\caption{Comparison between (a) LES-DEM simulation and (b) experiment. For the simulation, $\mu$ = 0.6, $\mu_r$ = 0 and $e$ = 0.1. Multimedia view (for the simulation).}
	\label{morphology6}
\end{figure}

In the case of unresolved simulations involving subaqueous bedload, the values of the  coefficients of sliding friction $\mu$, restitution $e$, and rolling friction $\mu_r$ are not tabulated, but the first two are known to vary within 0.1 and 0.9 \cite{Schmeeckle, Maurin, Liu, Pahtz_3}. In unresolved simulations, the water flow is not computed in the small separation between two approaching spheres (almost colliding), and, therefore, the effects of water (damping, for instance) shall be embedded in the friction and restitution coefficients. The same applies to other effects not directly captured in simulations. In the particular case of DEM simulations using spheres,  $\mu_r$ can be tuned in order to emulate angular grains \cite{Derakhshani} (the resistance to roll emulates the angular surface). We thus carried out simulations in which we varied $\mu$, $\mu_r$ and $e$, and followed the bedform along time. In all cases investigated, the initial pile was deformed into a barchan dune, similar to previous experiments \cite{Alvarez}. However, different from experiments (where the water flow was imposed in a less abrupt way by turning a pump on), the initial condition of simulations was a fully-developed flow that flattened the initial pile much faster than in the experiments.  Please see Figs. \ref{Setup} and \ref{morphology6} (Multimedia view) for movies showing the time evolution of numerically obtained barchans. After this initial transient, the dune behaved as in the experiments, and its characteristics varied with the grain properties. Based on measurements of celerity and morphology of dunes, we select the values of $\mu$, $\mu_r$ and $e$ for the proper LES-DEM simulations of subaqueous barchans, as shown next.

Figures \ref{morphology1} to \ref{morphology3} show snapshots of dunes (top view) for different coefficients of sliding friction $\mu$, rolling friction $\mu_r$ and restitution $e$, respectively, where each coefficient was varied independently of the others. In these figures, each column corresponds to the varied coefficient (values shown on the top), and each row to a time instant going from 0 to 20 s in steps of 2 s. We notice, as a general remark, that $\mu_r$ affects greatly the dune morphology, while the effect of $\mu$ is lesser, and that of $e$ even lesser.

\begin{figure}[h!]
	\centering
	\includegraphics[width=0.99\columnwidth]{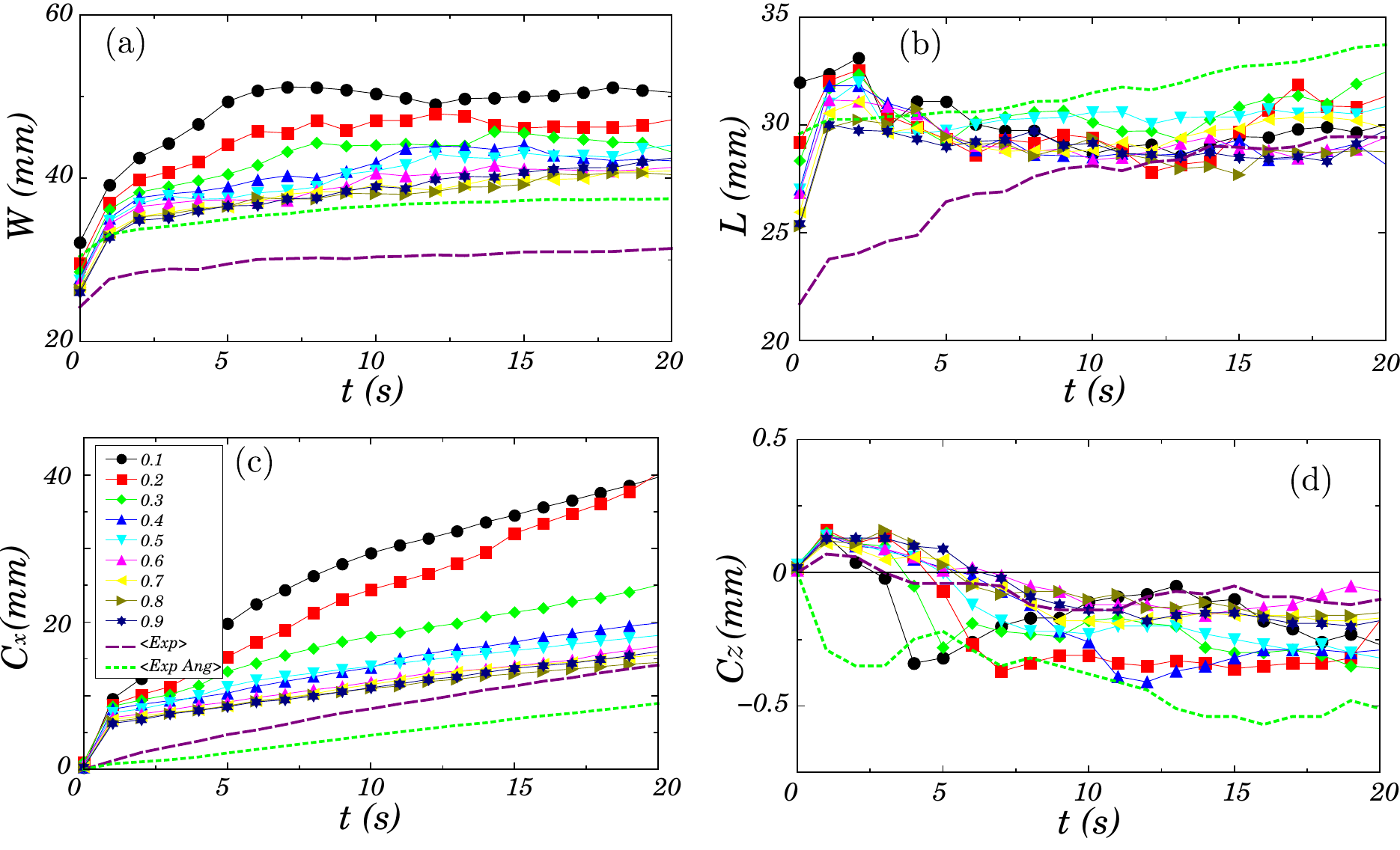}
	\caption{Time evolution of the morphology and positions of bedforms parameterized by the friction coefficient $\mu$: (a) Barchan width $W$; (b) barchan length $L$; (c) position of the centroid in the longitudinal direction $C_x$; and (d) position of the centroid in the transverse direction $C_z$. The symbols used for different values of $\mu$ are listed in the figure key, and $\mu_r$ = 0 and $e$ = 0.1 in the simulations.}
	\label{morphology4}
\end{figure}

From the snapshots in Fig. \ref{morphology1}, we observe larger differences for $\mu$ $\leq$ 0.4, both in terms of dune morphology and transport rate of grains: as $\mu$ decreases, the barchan becomes wider and larger quantities of grains are entrained further downstream from the horns (visible in the images). For $\mu$ $>$ 0.4, only small differences in the barchan shape are observed (investigated in detail in Fig. \ref{morphology4}). In its turn, Fig. \ref{morphology2} shows that changes in the restitution coefficient within 0.1 $\leq$ $e$ $\leq$ 0.9 have negligible effect on the morphology. Those observations are in agreement with previous computations of bedload using DEM \cite{Maurin, Pahtz_3}, which showed negligible variations in the transport rate for 0.01 $\leq$ $e$ $\leq$ 1 and $\mu$ $>$ 0.4.

The rolling friction is especially useful when angular grains are simulated as spherical particles, since the effects of angularities are embedded in $\mu_r$. This means that values of $\mu_r$ are rather large for angular grains simulated as spheres in DEM codes: while the rolling friction of glass spheres is virtually zero ($\mu_r$ = 0.01), that of typical sand is $\mu_r$ $\approx$ 0.3 (Ref. \cite{Derakhshani}). The computational costs of such procedure are much lesser than those of producing angular grains numerically. Figure \ref{morphology3} presents the effects of increasing $\mu_r$, and, therefore, the angularity of particles composing the bedform, going from 0 (perfect spheres) to 0.3 (typical sand). For the same water flow of previous simulations, we observe that the main effect of angularity, within the simulated time, is to hinder the motion of grains, increasing the time for the growth of barchan and reaching much longer timescales in the case of sand ($\mu_r$ = 0.3).

\begin{figure}[h!]
	\centering
	\includegraphics[width=0.99\columnwidth]{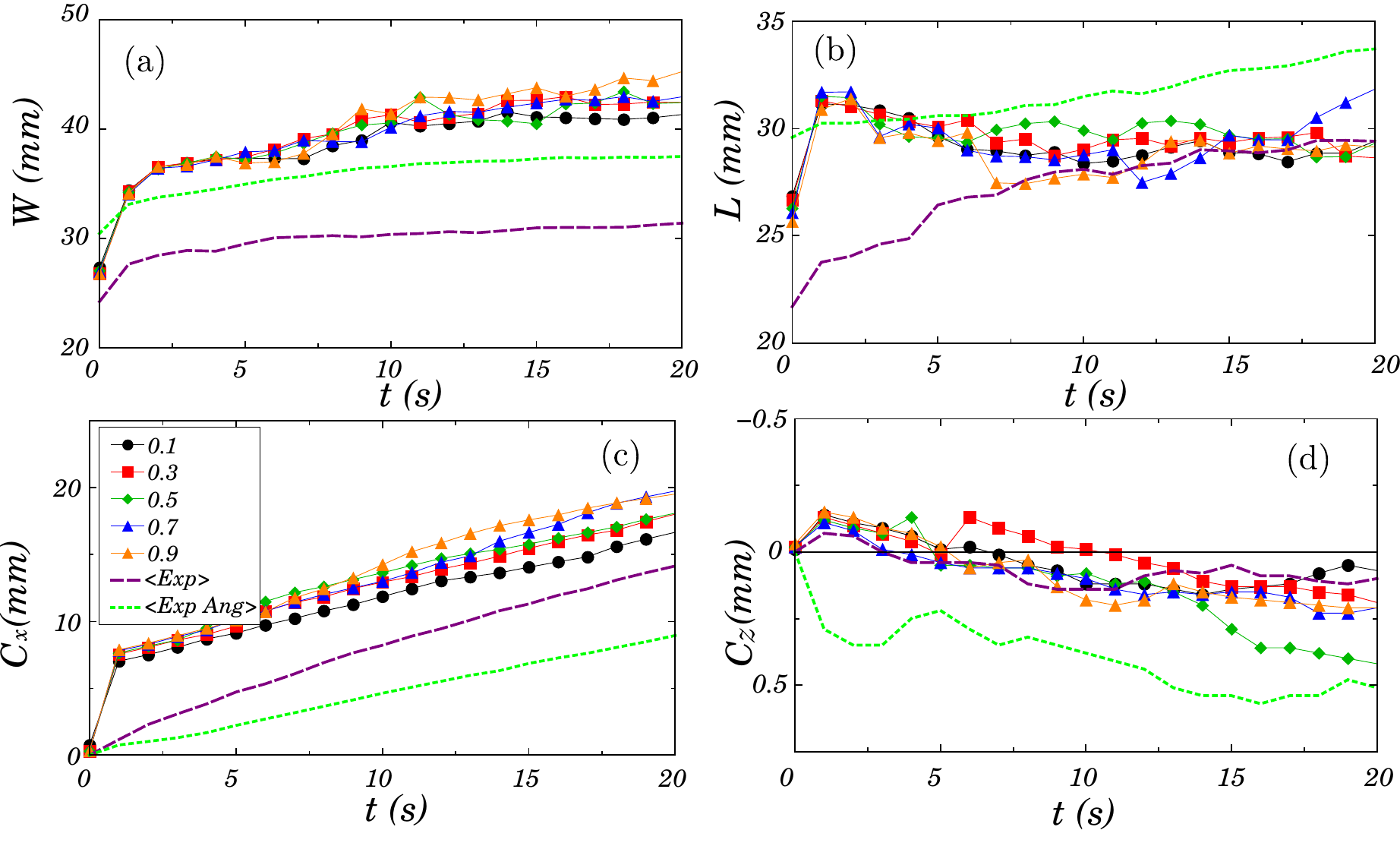}
	\caption{Time evolution of the morphology and positions of bedforms parameterized by the restitution coefficient $e$: (a) Barchan width $W$; (b) barchan length $L$; (c) position of the centroid in the longitudinal direction $C_x$; and (d) position of the centroid in the transverse direction $C_z$. The symbols used for different values of $e$ are listed in the figure key, and $\mu$ = 0.6 and $\mu_r$ = 0 in the simulations.}
	\label{morphology5}
\end{figure}

\begin{figure}[h!]
	\centering
	\includegraphics[width=0.99\columnwidth]{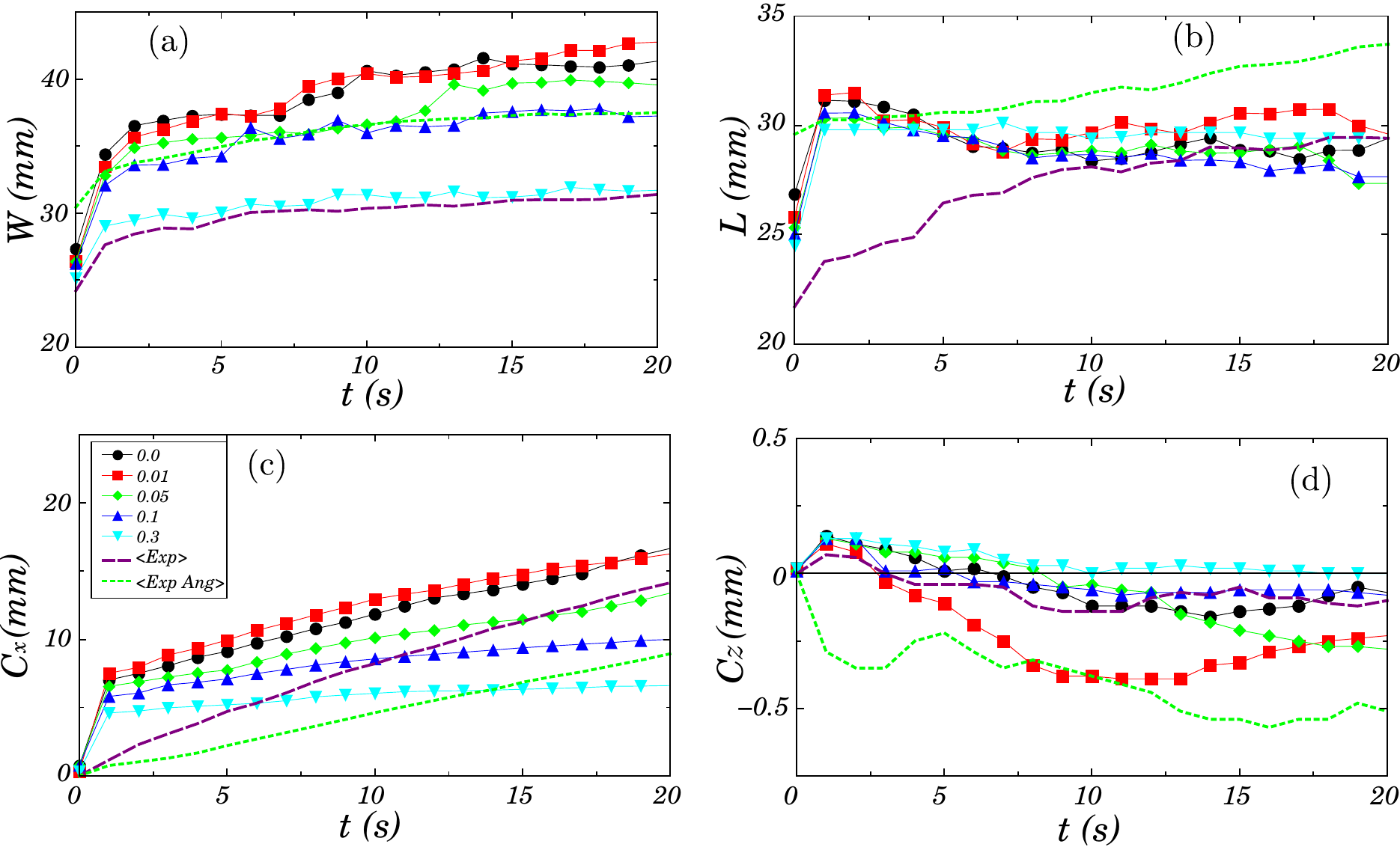}
	\caption{Time evolution of the morphology and positions of bedforms parameterized by the coefficient of rolling friction $\mu_r$: (a) Barchan width $W$; (b) barchan length $L$; (c) position of the centroid in the longitudinal direction $C_x$; and (d) position of the centroid in the transverse direction $C_z$. The symbols used for different values of $\mu_r$ are listed in the figure key, and $\mu$ = 0.6 and $e$ = 0.1 in the simulations.}
	\label{morphology7}
\end{figure}

In order to evaluate the friction coefficient $\mu$ for barchans consisting of glass spheres ($\mu_r$ $\approx$ 0), we plot in Fig. \ref{morphology4} the time evolution of the barchan width $W$ (Fig. \ref{morphology4}a), length $L$ (Fig. \ref{morphology4}b), longitudinal component of the centroid position $C_x$ (Fig. \ref{morphology4}c) and transverse component of the centroid position $C_z$ (Fig. \ref{morphology4}d) for different values of $\mu$ (the other coefficients were fixed), and compare them with experiments. The experiments were carried out as in Refs. \cite{Alvarez, Alvarez3, Alvarez4} (see the supplementary material for a brief description of the experimental setup), and each experiment shown in Fig. \ref{morphology4} corresponds to the average of three different test runs: one average for glass spheres ($<Exp>$ in the figure key) and the other for angular glass particles ($<Exp \; Ang>$ in the figure key). We observe a large deviation during the first few seconds of simulation due to the initial condition for the water flow mentioned previously: the initial flow condition in simulations is rather abrupt when compared with experiments, since the final realization of a developed channel flow (one-phase water flow simulated previously) is imposed as initial condition at $t$ = 0 s in the simulations with grains. The higher flow velocities at the very beginning of the simulations in comparison with those of experiments make the initial pile flatten faster in the numerical computations, displacing the $W$, $L$ and $C_x$ curves of simulations to higher values (0 s $<$ t $\lesssim$ 1 s). After this different transient, we observe that the numerical results present similar tendencies as those of the experiments, with closer results for 0.6 $\leq$ $\mu$ $\leq$ 0.9 (in agreement with Maurin et al. \cite{Maurin}, who found smaller variations of the transport rate of grains within 0.4 and 0.8). Because previous works computing  bedload with DEM used 0.6 \cite{Schmeeckle, Liu} or values close to it \cite{Kidanemariam, Kidanemariam3}, and since $\mu$ = 0.6 generates crescent shapes that are similar to those of experiments (Fig. \ref{morphology1}), we decided to fix $\mu$ = 0.6 for the simulations with glass spheres. For values lower than 0.5, dunes become wider and more grains are entrained further downstream.

We carried out similar evaluations for the coefficients of restitution $e$ and rolling friction $\mu_r$, which are shown in Figs. \ref{morphology5} and \ref{morphology7}, respectively, Figs. \ref{morphology5}a and Figs. \ref{morphology7}a to \ref{morphology5}d and \ref{morphology7}d corresponding to $W$, $L$, $C_x$ and $C_z$, respectively. Apart from the initial transient (different from the experiments because of the initial conditions for the water), the restitution coefficient has little effect on the results. This agrees with the results of Maurin et al. \cite {Maurin} and P\"ahtz et al. \cite{Pahtz_3} , who showed a very small influence of $e$ on CFD-DEM-computed bedload. In order to remain closer to values used in previous works \cite{Schmeeckle, Liu}, we fixed $e$ = 0.1. For varying $\mu_r$, we note that differences in $W$ and $L$ between simulations and experiments are less prominent for angular grains. Indeed, a reasonable agreement between experiments and simulations occurs for $\mu_r$ = 0.1 (for the angular particles that we used). However, this better agreement (in comparison with spherical particles) seems to be caused by relatively small effects of initial conditions on angular grains (due to their higher resistance to move). In general, by increasing $\mu_r$ the longitudinal celerity decreases (lower mobility) and the morphology ($W$ and $L$) has small deviations, with the exception of $\mu_r$ = 0.3. Nevertheless, $\mu_r$ = 0.3 seems excessively high for simulations emulating angular grains, the corresponding bedload layer presenting very low mobility in simulations. Figure \ref{morphology6} (Multimedia view) shows a comparison between our LES-DEM simulation with ($\mu$ = 0.6, $\mu_r$ = 0 and $e$ = 0.1) and one of our experiments for glass spheres. The agreement is good.

We observe a small deviation of $C_z$ toward negative values in Figs. \ref{morphology4}d, \ref{morphology5}d and \ref{morphology7}d. Since the physics is symmetric in the $z$ direction, $C_z$ should vary around a zero value. However, we used the same initial flow field (one-phase water flow), which had a small $z$ asymmetry, in the simulations with grains, so that the same initial asymmetry appeared in all simulations. We note that the asymmetries shown in Figs. \ref{morphology4}d, \ref{morphology5}d and \ref{morphology7}d are relatively small, being of the order of the grain diameter. This means that very few grains that spread away from the initial pile (at the beginning of simulations) can create such asymmetry.

Finally, we note that we kept two of the following coefficients at $\mu$ = 0.6, $\mu_r$ = 0.0 and $e$ = 0.1, while the third one varied within the values in Tab. \ref{diameterAndProps}, the fixed values being chosen in accordance with those found in the literature for simulations of subaqueous bedload \cite{Schmeeckle, Maurin, Liu, Pahtz_3}. Since the coefficients to be used for barchans (based on the results for their morphology) agree with those for subaqueous bedload, and given the large spans over which we varied the coefficients, the number of simulations carried out in this work seems reasonable.

\subsection{Dynamics at the grain scale}

\begin{figure}[h!]
	\centering
	\includegraphics[width=0.99\columnwidth]{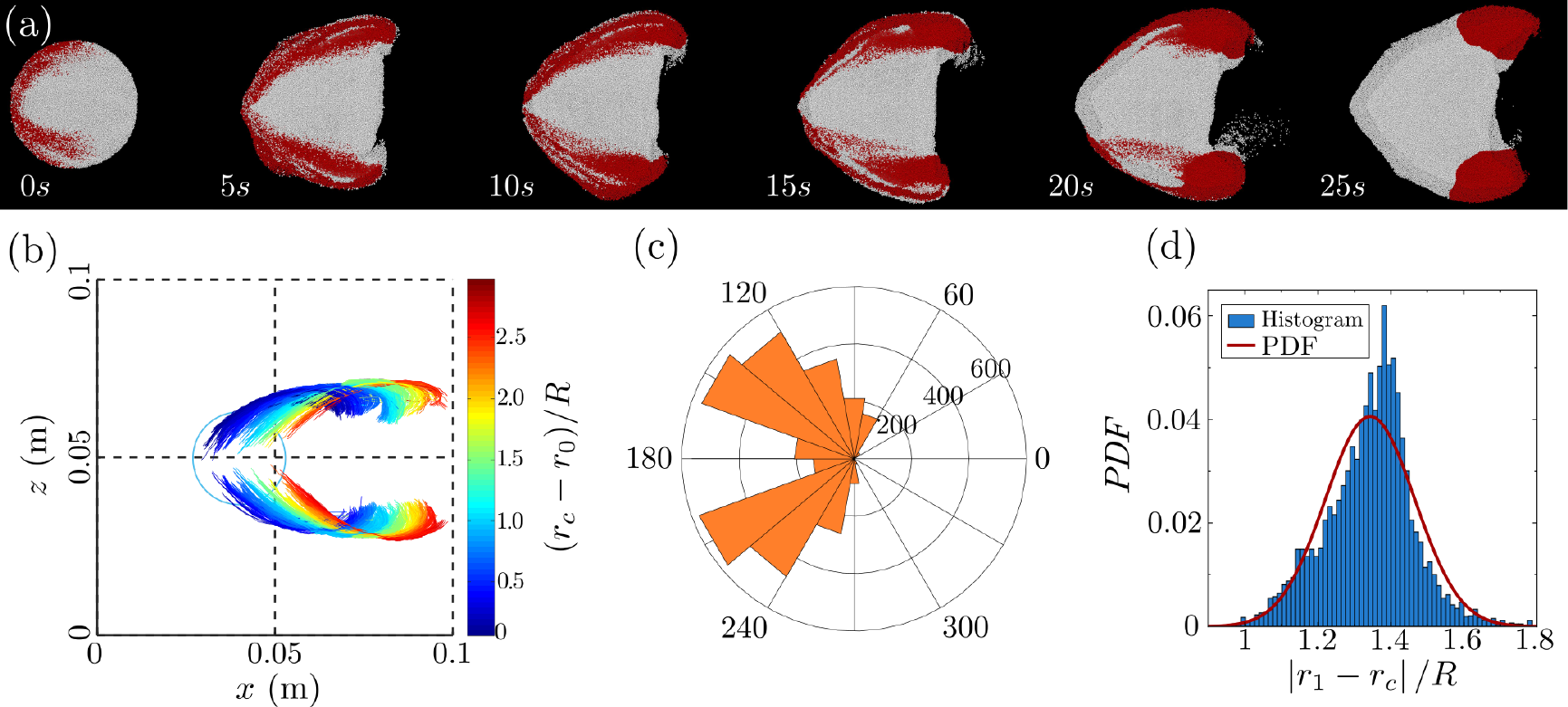}
	\caption{(a) Snapshots showing the deformation of the initial pile into a barchan dune, in which the particles that composed the horns at $t$ = 25 s are highlighted in red color since the beginning of the simulation. (b) Trajectories of the glass spheres that migrated to horns during the first $100$ s of simulation. The circle represents the initial pile and the color of pathlines varies with the instantaneous position of the dune centroid (as shown in the colorbar). (c-d) Frequencies of occurrence of the initial positions of the grains that migrated to horns during the first $100$ s as functions of (c) the angle with respect to the main flow direction and (d) the radial position. The direction of the water flow is $0^{\circ}$, and in figure (d), the probability density function (PDF) is also shown.  In the figures, $\mu$ = 0.6, $\mu_r$ = 0.0 and $e$ = 0.1.}
	\label{grain_dynamics}
\end{figure}

For the simulation with $\mu$ = 0.6, $\mu_r$ = 0.0 and $e$ = 0.1 (corresponding to glass spheres), we followed the grains of the initial pile as it was deformed into a barchan dune, and compared their trajectories with those of the experiments and numerical simulations of Alvarez and Franklin \cite{Alvarez3, Alvarez4, Alvarez5, Alvarez7}. We begin by tracking back, up to $t$ = 0 s, the grains that formed part of the horns when $t$ = 25 s. Figure \ref{grain_dynamics}a presents top view images of the bedform, in which the particles that formed the horns (or great part of them) at $t$ = 25 s are highlighted in red color since the beginning of the simulation. The figure thus depicts how the ensemble of grains migrates toward the horns as the initial pile is deformed into a barchan dune, and we observe that they come from the periphery of the upstream half of the initial pile. This is corroborated by Figs. \ref{grain_dynamics}c and \ref{grain_dynamics}d, which show the frequencies of occurrence of the initial positions of the grains that migrated to horns during the first $100$ s as functions of $\phi$ and radial position, respectively. From Figs. \ref{grain_dynamics}c and \ref{grain_dynamics}d, we notice that most of grains have their origin on the periphery of the bedform, within 105$^{\circ}$ $\leq\,\phi\,\leq$ 160$^{\circ}$ and 210$^{\circ}$ $\leq\,\phi\,\leq$ 260$^{\circ}$, where $\phi$ = $0^{\circ}$ in the flow direction. These results are in agreement with those of Refs. \cite{Alvarez3, Alvarez4, Alvarez5, Alvarez7}.

\begin{figure}[h!]
	\centering
	\includegraphics[width=0.95\columnwidth]{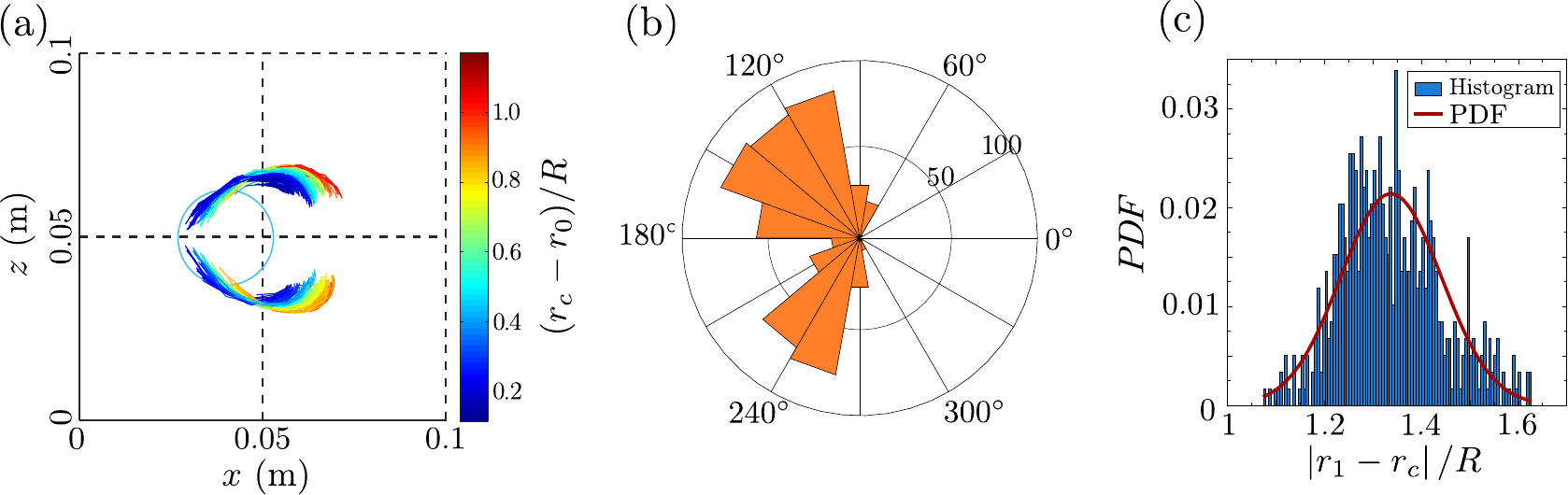}
	\caption{(a) Trajectories of the angular grains that migrated to horns during the first $50$ s of simulation. The circle represents the initial pile and the color of pathlines varies with the instantaneous position of the dune centroid (as shown in the colorbar). (b-c) Frequencies of occurrence of the initial positions of the grains that migrated to horns during the first $50$ s as functions of (b) the angle with respect to the main flow direction and (c) the radial position. The direction of the water flow is $0^{\circ}$, and in figure (c), the probability density function (PDF) is also shown.  In the figures, $\mu$ = 0.6, $\mu_r$ = 0.1 and $e$ = 0.1.}
	\label{grain_dynamics_angular}
\end{figure}

Figure \ref{grain_dynamics}b shows the trajectories of all grains that migrated to horns during the first $100$ s of simulation. In this figure, the circle represents the initial pile and the color of pathlines varies with the instantaneous position of the dune centroid, $\left( r_c - r_0 \right)/R$ (shown in the colorbar), where $r_c$ is the instantaneous position of the dune centroid, $r_0$ the position of the centroid of the initial pile, and $R$ the radius of the initial pile. Therefore, trajectories represented by dark-blue pathlines began when the pile was at the initial position and dark-red pathlines when the barchan had already migrated a distance greater than the pile diameter (2$R$). We observe that, as shown experimentally by Alvarez and Franklin \cite{Alvarez3, Alvarez4}, grains migrating to the horns of subaqueous barchans follow curved paths with considerable transverse displacements, different from the aeolian case, where grains are entrained by the fluid mainly in the longitudinal direction \cite{Bagnold_1} (with transverse movements occurring by reptation and avalanches). The same patterns occur for angular particles ($\mu_r$ = 0.1), as shown in Fig. \ref{grain_dynamics_angular}.

\begin{figure}[h!]
	\centering
	\includegraphics[width=0.9\columnwidth]{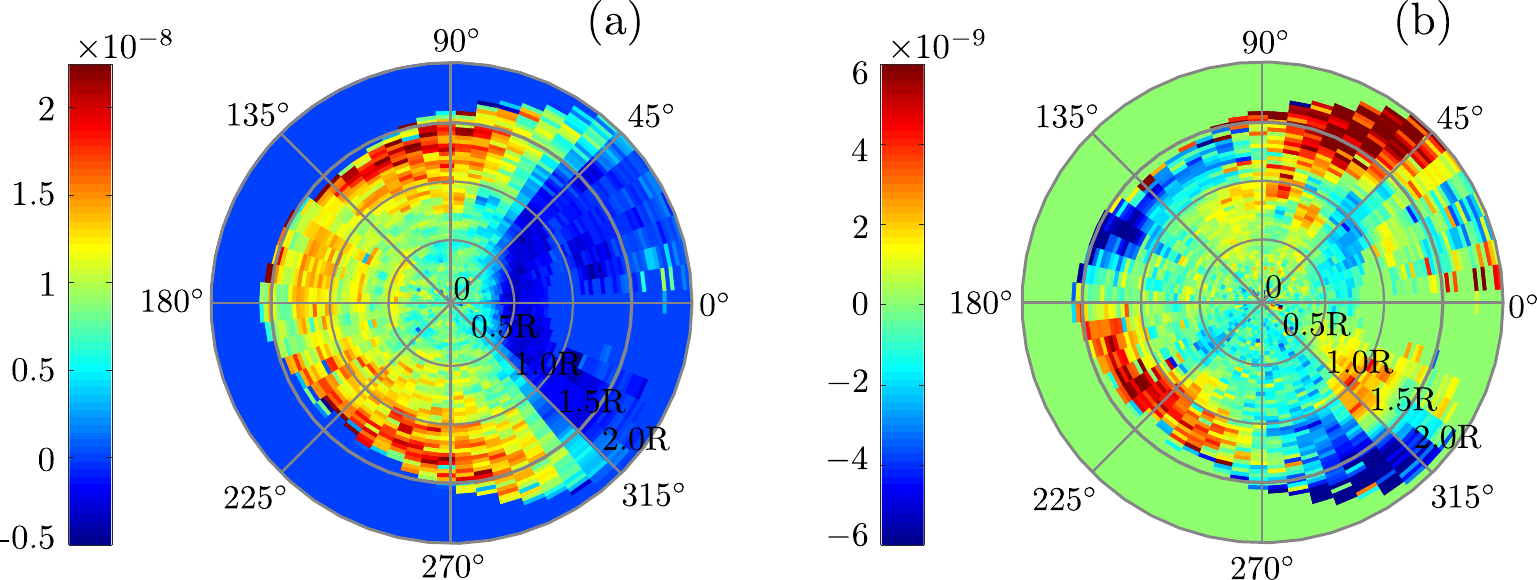}
	\caption{Spatial distributions of the (a) longitudinal and (b) transverse components of the ensemble-averaged resultant force acting on grains. The averages were computed within 10 s $\leq$ $t$ $\leq$ 20 s for all grains. The values shown in the colorbar are in N.}
	\label{grain_force}
\end{figure}

At each instant, the LES-DEM computations output the forces acting on individual grains. The summation of all forces (RHS of Eq. \ref{Fp}) provides thus the resultant force on each grain, something that until now is inaccessible from experiments. This is a great advantage of LES-DEM simulations over experiments and other numerical approaches.

Following the same procedure of Alvarez and Franklin \cite{Alvarez7}, we divided the dune into small regions and computed, along time, the ensemble average of the resultant force for all grains that passed by (or remained in) those regions, including grains below the dune surface. With that, we produced maps showing the spatial distributions of the resultant force acting on grains, such as in Fig. \ref{grain_force}. Figures \ref{grain_force}a and \ref{grain_force}b present, respectively, spatial distributions of the longitudinal ($F_x$) and transverse ($F_z$) components of the ensemble-averaged force, computed within 10 s $\leq$ $t$ $\leq$ 20 s, in polar coordinates (the water flow direction is 0$^{\circ}$). In these figures, $F_x$ is positive toward downstream ($\phi$ = 0$^{\circ}$) and negative toward upstream ($\phi$ = 180$^{\circ}$), while $F_z$ is positive toward the right (with respect to the flow direction, i.e., $\phi$ = 270$^{\circ}$) and negative toward the left direction ($\phi$ = 90$^{\circ}$). Since there are no experimental measurements of the resultant force acting on each grain, we compare the present results directly with our previous simulations and indirectly, in terms of expected trajectories, with previous experiments.

As in Ref. \cite{Alvarez7}, we observe that the longitudinal component is positive upstream the crest and negative on the crest, indicating a deceleration in that latter region and, consequently, grain deposition on the crest. In the inter-horns region downstream the crest, we observe negative values caused by the recirculation bubble, showing that grains in that region are entrained upstream and kept close to the lee face. We also observe that the longitudinal component is stronger close to the lateral flanks of the barchan 100$^{\circ}$ $\lesssim\,\phi\,\lesssim$ 160$^{\circ}$ and 210$^{\circ}$ $\lesssim\,\phi\,\lesssim$ 260$^{\circ}$. On the dune periphery, we observe that the transverse component points outwards upstream the crest and inwards downstream the crest. The force distribution agrees with previous experiments on subaqueous barchans \cite{Alvarez3, Alvarez4}, which showed that grains migrating to horns come from upstream regions on the periphery of the dune, executing curved trajectories with considerable transverse motion.

When LES-DEM computations with $\mu$ = 0.6, $\mu_r$ = 0.0 and $e$ = 0.1 are used for simulating subaqueous barchans, the results at the grain scale agree well with experiments under similar conditions (glass spheres) \cite{Alvarez3, Alvarez4}. This type of simulation can be further explored to investigate dunes that consist of grains of different materials by using the proper coefficients (for example, $\mu_r$ = 0.3 for sand \cite{Derakhshani}).

Finally, we observed that the best results are obtained when the ratio between the sizes of the smallest mesh and mean particle diameter is within 1.2 and 1.6. Outside this range, grains tend to accumulate excessively in upstream regions, the dune shape deviates from experimental results, or the dune motion is different from experimentally observed. We note that, even though our simulations present an excellent agreement with experiments, not all the details are well captured, such as the small jumps of some particles observed in the experiments. In our opinion, resolved computations (including angularities in the case of angular particles) should be employed to capture those small details, but in that case simulations would be excessively time consuming. Unresolved LES-DEM computations bear a good compromise between the accuracy of results and computational costs for the numerical simulation of barchan dunes.

\section{CONCLUSIONS}
\label{sec:conclusions}

In this paper, we inquired into the values of solid-solid parameters, namely the coefficients of sliding friction $\mu$, rolling friction $\mu_r$ and restitution $e$, for: (i) carrying out CFD-DEM computations that successfully reproduce subaqueous barchans (as measured in previous and current experiments); and (ii) understanding the influence of those parameters on the barchan dynamics. We made use of the open-source code \mbox{CFDEM}, which couples the open-source codes OpenFOAM (for CFD) with LIGGGHTS (for DEM), and we computed the fluid flow using LES in order capture the most energetic vortices (in particular in the recirculation region). We showed, for the first time, the ranges of parameters ($\mu$, $\mu_r$ and $e$) for the proper computation of barchans down to the grain scale, obtaining the dune morphology and timescales, disturbances of the fluid flow, trajectories of individual grains, and forces experienced by each grain. Based on experiments in the subaqueous case, we found that $\mu$ = 0.6, $\mu_r$ = 0.0 and $e$ = 0.1 are appropriate for bedforms consisting of glass spheres, and $\mu$ = 0.6, $\mu_r$ = 0.1 and $e$ = 0.1 for bedforms of angular glass particles. These values have been determined by varying the coefficients and comparing the results directly with experiments. In addition, we observed that the best results are obtained when the ratio between the sizes of the smallest mesh and mean particle diameter is within 1.2 and 1.6. Finally, we made a parametric study and showed how different grain properties affect the morphology of barchans. Our results show how to configure LES-DEM computations to resolve, at the same time, the morphology of barchans, the grains' motion, and details of the fluid flow, opening new possibilities for studying the dynamics of dunes down to the grain scale (including forces on each grain, which are not accessible from current experiments).

\section*{AUTHOR DECLARATIONS}
\noindent \textbf{Conflict of Interest}

The authors have no conflicts to disclose

\section*{SUPPLEMENTARY MATERIAL}
See the supplementary material for a brief description of the experimental setup and additional figures and graphics of our results.

\section*{DATA AVAILABILITY}
The data that support the findings of this study are openly available in Mendeley Data at https://data.mendeley.com/datasets/zhnngnvvz8.

\begin{acknowledgments}
The authors are grateful to FAPESP (Grant Nos. 2018/14981-7,  2019/10239-7 and 2019/20888-2) for the financial support provided. Carlos A. Alvarez is grateful to SENESCYT for admitting his stay as a postdoctoral researcher outside Ecuador.
\end{acknowledgments}

\appendix
\section{Contact model}
\label{appendix}

When two spherical particles $i$ and $j$ are in contact, their normal displacement $\delta_{n}$ is given by:

\begin{equation}
	\delta_{n} = r_{i} + r_{j} - |\vec{x}_{i}-\vec{x}_{j}| \, ,
	\label{eqndeltan}
\end{equation}

\noindent where $\vec{x}_{i}$ and $\vec{x}_{j}$ are the positions of the centers of the particles, and $r_{i}$ and $r_{j}$ their radius. These particles have Young moduli $E_{i}$ and $E_{j}$ and Poisson's ratios $\sigma_{i}$ and $\sigma_{j}$. The coefficients $\kappa_{n}$, $\kappa_{t}$, $\gamma_{n}$, and $\gamma_{t}$ are computed by Eqs. \ref{eqnkappan} to \ref{eqntgamma}, which are based on the displacement $\delta_{n}$, damping coefficient $\beta$ (Eq. \ref{eqnbeta}), and the effective radius $r_{c}$, mass $m_{c}$, contact modulus $E_{c}$ and shear modulus $G_{c}$ (Eqs. \ref{eqnrc} to \ref{eqnGc}).

\begin{equation}
	r_{c} = \frac{r_{i}r_{j}}{r_{i}+r_{j}}
	\label{eqnrc}
\end{equation}

\begin{equation}
	m_{c} = \frac{m_{i}m_{j}}{m_{i}+m_{j}}
	\label{eqnmc}
\end{equation}

\begin{equation}
	E_{c} = \Bigg(\frac{1 - \sigma_{i}^{2}}{E_{i}} + \frac{1 - \sigma_{j}^{2}}{E_{j}}\Bigg)^{-1}
	\label{eqnEc}
\end{equation}

\begin{equation}
	G_{c} = \Bigg[\frac{2(2 - \sigma_{i})(1 + \sigma_{i})}{E_{i}} + \frac{2(2 - \sigma_{j})(1 + \sigma_{j})}{E_{j}}\Bigg]^{-1}
	\label{eqnGc}
\end{equation}

\begin{equation}
	\kappa_n = \frac{4}{3} E_c \sqrt{R_c\delta_n}
	\label{eqnkappan}
\end{equation}

\begin{equation}
	\kappa_t = 8 G_c \sqrt{R_c\delta_n}
	\label{eqnkappat}
\end{equation}

\begin{equation}
	\beta = \frac{\ln(\epsilon)}{\sqrt{\ln^{2}(\epsilon) + \pi^2}}
	\label{eqnbeta}
\end{equation}

\begin{equation}
	\gamma_n = -2\sqrt{\frac{5}{6}} \beta \sqrt{2E_c \sqrt{R_c\delta_n} m_c}
	\label{eqngamman}
\end{equation}

\begin{equation}
	\gamma_t = -2\sqrt{\frac{5}{6}} \beta \sqrt{8 G_c \sqrt{R_c\delta_n} m_c}
	\label{eqntgamma}
\end{equation}

The contact torques are computed as the sum of the torques due to $F_{c,t}$ and rolling friction $\vec{T}_r$. For a given particle of radius $r$, contact torques are given by Eq. \ref{eqntorques}:

\begin{equation}
	\vec{T}_{c} = \sum \left( r F_{c,t}\vec{n} \times \vec{t} + \vec{T}_{r}\right) \, ,
	\label{eqntorques}
\end{equation}

\noindent where $\vec{n}$ and $\vec{t}$ are unit vectors in the normal and tangential directions, respectively. Derakhshani et al. \cite{Derakhshani} showed that $\vec{T}_r$ can be modeled by considering only the spring component (the damping one being neglected). For a rolling stiffness $k_r$ (Eq. \ref{eqnrollingstiff}) and an incremental rolling $\theta_r$ (at the considered contact), 

\begin{equation}
	\vec{T}_{r} = -k_r \Delta \theta_r \vec{n} \times \vec{t} \, ,
	\label{eqntorquerolling}
\end{equation}

\begin{equation}
	k_r = \mu_r R_c \frac{F_{c,n}}{\theta_r^m} \, ,
	\label{eqnrollingstiff}
\end{equation}

\noindent where $\theta_r^m$ is the angle for incipient rolling and $\mu_r$ is the coefficient of rolling resistance.

\section{CFD-DEM force exchange}
\label{appendix2}

This section describes how the fluid-particle interactions are computed for the DEM ($\vec{F}_{fp}$, Eq. \ref{Ffp_sim}) and CFD ($\vec{f}_{exch}$, Eq. \ref{forces_exchange}) parts.

\subsection{CFD}

The drag force in fluid equation, corresponding to the first term on the RHS of Eq. \ref{forces_exchange}, is computed by the Gidaspow model \cite{Gidaspow},

\begin{equation}
	{\vec{f}_{d}=\frac{1}{\Delta V}\sum_{i}^{n_{p}}\frac{V_{p}\beta_G}{1-\alpha_{f}}\left(\vec{u}_{p}-\vec{u}_{fp}\right)} \, ,
	\label{gidaspow}
\end{equation}

\noindent where $\vec{u}_{fp}$ is the fluid velocity at the particle position, and

\begin{equation}
	\beta_G=\left\{
	\begin{array}{cc}
		\frac{3}{4}C_{d}\frac{\rho_{f}\alpha_{f}\left(1-\alpha_{f}\right)\left|\vec{u}_{f}-\vec{u}_{p}\right|}{d}\alpha_{f}^{-2.65} & \left(1-\alpha{f}\right) \leq 0.2\\
		\\
		150\frac{\left(\alpha_{f}\right)^{2}\vec{u}_{f}}{\alpha_{f}d^{2}}+1.75\frac{\rho_{f}\left(1-\alpha_{f}\right)\left|\vec{u}_{f}-\vec{u}_{p}\right|}{d} & \left(1-\alpha{f}\right) > 0.2\\
	\end{array}\right.
	\label{beta}
\end{equation}

\noindent with,

\begin{equation}
	C_{d}=\left(0.63+0.48\sqrt{V_{r}/Re_d}\right) \, ,
	\label{cd}
\end{equation}

\noindent $Re_d$ being the particle Reynolds number, given by

\begin{equation}
	Re=\frac{\rho_{f}d\left|\vec{u}_{f}-\vec{u}_{p}\right|}{\mu_{f}} \, , 
	\label{Re}
\end{equation}

\noindent and $V_{r}$ is given Eq. \ref{Vr},

\begin{equation}
	V_{r}=0.5\left(A_{1}-0.06Re+\sqrt{\left(0.06\right)^{2}+0.12Re\left(2A_{2}-A_{1}+A_{1}^{2}\right)}\right) \, ,
	\label{Vr}
\end{equation}

\noindent with,

\begin{equation}
	\begin{array}{cc}
		A_{1}=\alpha_{f}^{4.14} \\
		\\
		A_{2}=\left\{
		\begin{array}{cc}
			0.8\alpha_{f}^{1.28} & \alpha_{f} \leq 0.85\\
			\\
			\alpha_{f}^{2.65} & \alpha_{f} > 0.85
		\end{array}\right.
	\end{array}
	\label{A}
\end{equation}

The second term on the RHS of Eq. \ref{forces_exchange} corresponds to the virtual mass force by unit of volume, $\vec{f}_{vm}$, given by Eq. \ref{Fvm}.

\begin{equation}
	\vec{f}_{vm}=\frac{1}{2}\left(1-\alpha_{f}\right)\rho_{f}\left(\frac{d\vec{u}_{f}}{dt}-\frac{d\vec{u}_{p}}{dt}\right)
	\label{Fvm}
\end{equation}

The exchange term in the fluid equation (Eq. \ref{momentum_fluid}) is thus $\vec{f}_{exch}$ = $\vec{f}_{d}$ + $\vec{f}_{vm}$.

\subsection{DEM}

For each grain of volume $V_p$, the forces in Eq. \ref{Ffp_sim} are computed by Eqs. \ref{Fd_appendix} to \ref{Fvm_appendix}.

\begin{equation}
	\vec{F}_d = V_p \vec{f}_{d}
		\label{Fd_appendix}
\end{equation}

\begin{equation}
	\vec{F}_p = -V_p\nabla P
	\label{Fp_appendix}
\end{equation}

\begin{equation}
	\vec{F}_{\tau} = V_p \nabla\cdot\vec{\vec{\tau}}
	\label{Ft_appendix}
\end{equation}

\begin{equation}
	\vec{F}_{vm} = \frac{1}{2} V_p \rho_{f}\left(\frac{d\vec{u}_{f}}{dt}-\frac{d\vec{u}_{p}}{dt}\right)
	\label{Fvm_appendix}
\end{equation}

\section{Smoothing of the void fraction and momentum}
\label{appendix3}

In order to make computations of coarse grains (grain size comparable to that of meshes) more stable, a difusion equation can be used for the void fraction as well as the coupling forces:

\begin{equation}
	\frac{\partial \xi}{\partial t} = \frac{\lambda}{\Delta t_{CFD}}\nabla^2\xi
	\label{diffusion_appendix}
\end{equation}

\noindent where $\lambda$ is a characteristic smoothing length, $\Delta t_{CFD}$ is the time step of CFD computations, and $\xi$ = $\alpha_f$ or $\vec{f}_{exch}$. In the present simulations, we considered $\lambda$ = 3$d$.

\bibliography{references}

\end{document}